\pdfoutput=1
\documentclass[aps,pra,reprint,showpacs]{revtex4-1}
\usepackage{graphicx}
\usepackage{amsmath}
\usepackage{amssymb}
\usepackage{amsfonts}
\usepackage{dcolumn}
\usepackage{dsfont}
\usepackage{latexsym}
\usepackage{rotating}
\usepackage{color}
\usepackage{latexsym}
\usepackage{bbm}
\usepackage{subfigure}
\usepackage{float}
\usepackage{epsfig}
\usepackage{psfrag}
\usepackage{natbib}
\usepackage{bm}
\usepackage{amsthm}
\usepackage{eucal}
\usepackage{mathrsfs}
\usepackage{url}
\usepackage{braket}
\usepackage{mathtools}
\usepackage{amsmath,amssymb}
\usepackage{epsfig}
\usepackage{amsmath,amssymb}
\usepackage{hyperref}

\usepackage{color}

\usepackage{hyperref}
\hypersetup{
colorlinks=true,final=true,
        linkcolor=blue,
        citecolor=blue,
        filecolor=blue,
        urlcolor=blue,
}

\begin{document}

\title{Dynamics of Airy pulse under phase modulation: Few interesting aspects}
\author{Aritra Banerjee$^\star$ and Samudra Roy$^\dagger$}
\affiliation{Department of Physics, Indian Institute of Technology Kharagpur, W.B. 721302, India}

\email{$^\dagger$samudra.roy@phy.iitkgp.ernet.in\\
$^\star$aritra@iitkgp.ac.in}

\begin{abstract}

In this work we have tried to study the interesting dynamics of Airy pulse under phase modulation. We investigate the role of linear, quadratic and cubic phase modulation  on Airy pulse dynamics in pure linear regime.  As a specific example, we study the influence of cubic phase modulation (CPM) as a self healing effect when the propagation dynamics of a finite energy Airy pulse is perturbed under the third order dispersion (TOD). As a consequence of TOD, the pulse flips in time domain and propagates with a reverse acceleration. This unusual propagation characteristic can be restricted through CPM which counterbalance the TOD induced flipping phenomenon. With proper analytical and numerical treatment we demonstrate, how CPM not only resists the temporal flipping but helps in preserving the pulse shape under perturbation. We put special emphasis on the study of the Airy pulse at flipping point where it loses its characteristic shape  and converts itself to a pure Gaussian pulse. We derive the Gaussian width analytically and demonstrate that a suitable quadratic phase modulation can focus the Airy pulse to a point and leads to a chirp-free Gaussian pulse. The present study is useful in understanding the physical insight of the unique Airy dynamic under phase modulation.      

\end{abstract}

\maketitle

\section{Introduction}
In a famous paper in 1979 Berry and Balazs, first introduced the idea of the infinite energy Airy wave packet in the context of quantum mechanics \cite{Berry}. They have shown that this wave packet satisfies the Schr\"{o}dinger’s equation in free space, so it moves without any distortion. They have also shown that it has a unique feature of following a parabolic trajectory due to the transverse acceleration of the wave packet which was later confirmed by D. Greenberger based on the theory of equivalence principle    \cite{Daniel}. In 2007, Siviloglou and Christodoulides \cite{Siviloglou} exploited the model of Airy wave packet by introducing a finite energy Airy beam (FEAB) in the optical domain which can resist diffraction up to a finite distance and bend in space without any potential. This free acceleration of the Airy packet was also observed experimentally  \cite{Siviloglou_b}. Since this remarkable breakthrough, many works have been done on exploiting the unique properties of the FEAB like, self acceleration, quasi diffraction free and self-healing nature \cite{Siviloglou,Siviloglou_b,Broky}. The optical  behaviour of the FEAB in nonlinear regime is also explored. Chen et. al. demonstrated the effect of Kerr nonlinearity on Airy beam  where compression of major Airy lobe was observed \cite{Rui}. The existence of nonlinear Airy like solutions in presence of third order Kerr nonlinearity and nonlinear losses have been reported subsequently \cite{Lotti,Lotti_b}.  Because of these unique properties in linear and nonlinear regime, FEAB has been useful in many applications such as curved filament generations \cite{polynkin,polynkin_b}, abrupt auto focussing \cite{Couairon}, manipulation of nanoparticles using FEAB\cite{Dholakia,P. Zhang}, optical routing \cite{Rose}, tailored femtosecond filament generation \cite{Papazoglou} etc.
Exploiting the isomorphism between the spatial diffraction and the temporal dispersion, the concept of  finite energy Airy pulse (FEAP) is introduced recently which is a temporal analogue of Airy beam \cite{Saari,Ament}. The FEAB bends in space due to its acceleration in the transverse direction where as the acceleration of the temporal Airy pulse is characterized by its varying group velocity. This property seeks great interest for the problems of the dynamics of the Airy pulses for both, in linear and nonlinear regimes. In linear regime many works have already been reported where the role of dominant  third order dispersion (TOD) on the Airy pulse is investigated \cite{driben,Shaarawi}. The dynamics of the FEAP in nonlinear regime has also been studied extensively. Soliton formation from Airy pulse \cite{Fattal} and its control through quadratic phase modulation \cite{Lifu}, effect of Raman scattering, self steeping and Kerr nonlinearity \cite{Chen,Hu}, supercontunuum generation \cite{Ament}, periodic dispersion modulation \cite{Wang}, effect of modulation instability \cite{L.Zhang} etc. are few  important works that have been reported recently in the context of FEAP. The description of Airy functions in time domain opens up exciting applications ranging from bioimaging, nano-machining to plasma physics \cite{Courvoisier,Englert,Gotte,javier,sarpe,Thomas}. In all these applications shape preservation of the FEAP is important.  In this work, we try to investigate the role of  phase modulation (PM) on controlling the shape of Airy pulse under the perturbation of TOD. We unfold some interesting aspects of pulse reshaping when PM is applied to the original pulse. We have restricted our study upto cubic phase modulation (CPM) in linear regime where we have already observed some interesting phenomenon which are less addressed.  We emphasise more in understanding the underlying physics of Airy pulse dynamics. In the pure linear regime, when the FEAP is launched very close to zero dispersion frequency (with dominant TOD), it changes its path abruptly with a reverse acceleration.  This flipping phenomenon of FEAP has already been reported by Driben et. al. \cite{driben}. In the flipping region the FEAP loses its shape and try to focus to a point. The size of this focusing area essentially depends on the relative strength of TOD. The temporal  reversibility of the Airy pulse is an inevitable phenomenon inside a dispersive waveguide where TOD coefficient is positive. In this report, we have proposed a recipe which can prevent this temporal flipping and preserve the pulse shape for relatively large distance. Numerically and analytically we show that, a suitable PM of the input spectra of the FEAP can control the rate of acceleration of the pulse and hence the flipping point can be tailored. The role of quadratic phase modulation (QPM) in the context of manipulation of the soliton emitted from FEAP was studied earlier \cite{Lifu}. However, to the best of our knowledge, the influence of  phase modulation on the dynamics of a perturbed Airy pulse in linear regime is unexplored. In this work we have shown that QPM can squeeze the FEAP to a point (we call it as absolute focusing).  At absolute focusing the shape of the Airy pulse is lost completely and we have a pure chirp-free Gaussian pulse.

We organize this work as follows: In section II, we introduce the governing equation of a  Airy pulse moving in a dispersive medium.  We solve the equation both analytically and numerically. The closed form solution of a general phase modulated FEAP is also introduced in this section. In section III we discuss our results which is divided into three subsections where the role of linear, quadratic, cubic  phase modulation on a perturbed Airy pulse is described elaborately. We analytically demonstrate that, CPM is the only modulation which helps in preserving the shape of a  FEAP perturbed by TOD for a larger propagation distance. We have put special emphasis in understanding the behaviour of the Airy pulse at flipping point where it converted to a pure Gaussian pulse. We derive the characteristic Gaussian width and also propose a recipe to manipulate it via QPM. The flipping region is characterized by the width of the Gaussian pulse and one can even reduce this region in a point with proper QPM.  To confirm our findings  we solve the governing equation numerically, by adopting the standard split-step Fourier method \cite{Agarwal}. The simulation agrees well with analytical results. Finally, in section IV we summarize our findings.

\section{Propagation Model}
In the linear regime, with the perturbation of TOD, the dynamics of the amplitude of FEAP, $u(z,T)$, is governed by following partial differential equation \cite{Agarwal},

\begin{equation}\label{q1} 
i\frac{\partial u}{\partial z}=\frac{{{\beta }_{2}}}{2}\frac{{{\partial }^{2~}}u}{\partial {{T}^{2}}}+i\frac{{{\beta }_{3}}}{6}\frac{{{\partial }^{3~}}u}{\partial {{T}^{3}}},
\end{equation}
where ${\beta_2}$ and ${\beta_3}$ are the coefficients of $2^{nd}$ and $3^{rd}$ order dispersion which are obtained from Taylor's series expansion of the propagation constant ${\beta (\omega)}$. $z$ and $T$ are respectively, the space and time variable in the frame, that is moving with a group velocity $v_g$ . The normalised form of Eq. \eqref{q1} can be written as, 

\begin{equation} \label{q2}
i\frac{\partial U}{\partial \xi}=\frac{{{sgn(\beta }_{2})}}{2}\frac{{{\partial }^{2~}}U}{\partial {{\tau}^{2}}}+i{{{\delta }_{3}}}\frac{{{\partial }^{3~}}U}{\partial {{\tau}^{3}}}.
\end{equation}

Here the parameters are rescaled as,  $u=\sqrt{P_0}U$, $\xi=z{L_D}^{-1}$, $\tau=(t-z{v_g}^{-1})/t_0=T/{t_0}$. $P_0$, $t_0$ and $L_D=t_0^2|\beta_2|^{-1}$ are the input peak power, some initial pulse width and dispersion length respectively. The TOD parameter ${\beta_3}$ is rescaled as, $\delta_3=\beta_3/(3!|\beta_2|t_0)$. In absence of TOD ($\delta_3=0$) we have the nondispersive solution of Eq. \eqref{q2} as, $U(\xi,\tau)=Ai(\tau-\xi^2/4)\exp[i(\xi\tau/2-\xi^3/12)]$ in anomalous dispersion regime. At input, the pulse peaks at $\tau_{p0}\approx-(3\pi/8)^{2/3}$. Clearly the peak position ($\tau_p$) of the pulse shifts with the propagation distance as, $\tau_p\approx \tau_{p0}+\xi^2/4$. The $\xi^2/4$ term describes the ballistic trajectory  of an Airy pulse in the medium. Equation \eqref{q2} can be solved by using the standard Fourier transform  method. The general solution is given as,    
\begin{align} \label{q3}
U\left( \xi ,\tau  \right)=\frac{1}{2\pi}\underset{-\infty }{\overset{\infty }{\mathop \int }}\,\tilde{U}\left( 0,\omega  \right)\exp \left( i\frac{sgn({\beta_2}){{\omega }^2}\xi }{2}-i{{\delta }_{3}}{{\omega }^{3}}\xi  \right)\nonumber\\ \times\exp \left( i\omega \tau  \right)d\omega,
\end{align}
where $\tilde{U}(0,\omega)$ is the Fourier transform of the input field given by $U\left( 0,\tau  \right)=\sqrt{P_0}Ai(\tau)\exp(a\tau)$. The truncation parameter $a$, exponentially truncates the infinite energy pulse to the finite energy pulse. If we perform the Fourier transformation on this initial pulse in normalised frequency domain, we will get the initial spectrum $\tilde{U}(0,\omega)$ as,
\begin{equation} \label{q4}
\tilde{U}\left( 0,\omega  \right)=\exp \left( -a{{\omega }^{2}} \right)\exp i\left( \frac{{\omega ^3}}{3}-a^2\omega -\frac{i{{a}^{3}}}{3} \right).
\end{equation}
In our study, we have modulated this input spectra by a general phase modulation. Under a general phase modulation, the input spectra turns out to be,  
\begin{equation}\label{q5}
\tilde{U}\left( 0,\omega  \right)=\exp \left( -a{{\omega }^{2}}+i{{\mu}_{k}}{{\omega }^{k}} \right)\exp i\left( \frac{{{\omega }^{3}}}{3}-{{a}^{2}}\omega -~\frac{i{{a}^{3}}}{3} \right).
\end{equation}
Here the index $k=1,2,3...$ which stands for the degree of the modulation. The parameter ${\mu_k}$ represents the coefficient of the $k^{th}$ order phase modulation. It should be noted that, by plugging Eq. \eqref{q5} into Eq. \eqref{q3}, it is always possible to extract the analytical form of the output pulse which is initially phase modulated.

\section{Results and Discussions}

In this section we will study separately the  role of linear ; quadratic and cubic phase modulation  on the dynamics of the pulse in presence of TOD. In absence of any phase modulation (${\mu_k}$=0), the general solution of Eq. \eqref{q2} is, 

\begin{equation} \label{q6}
U\left( \xi ,\tau  \right)=\frac{1}{c}\exp \left( \frac{{{a}^{3}}}{3} \right)Ai\left( \frac{b}{c}-\frac{{{n}^{2}}}{{{c}^{4}}} \right)\exp\left( i\frac{2{{n}^{3}}}{3{{c}^{6}}}-i\frac{nb}{{{c}^{3}}} \right) ,
\end{equation}
where, $c={{\left( 1-3{{\delta }_{3}}\xi  \right)}^{\frac{1}{3}}}$; $n=[ ia+sgn(\beta_2)\xi/2]$ and $b=( \tau -{{a}^{2}})$. From the expression it is evident that the presence of positive TOD coefficient ($\delta_3>0$) leads to a singularity when $c=0$. The distance at which $c$ vanishes is termed as, the flipping point ($\xi_{flip}$).  The point $\xi_{flip}$ has significant importance since at this point the Airy pulse loses its characteristic shape and  inverted temporally. The expression of $\xi_{flip}$ simply comes out to be, 
\begin{equation} \label{q7}
{{\xi }_{flip}}=\frac{1}{3{{\delta }_{3}}}. 
\end{equation}

\begin{figure}[h!]
\begin{center} 

\epsfig{file=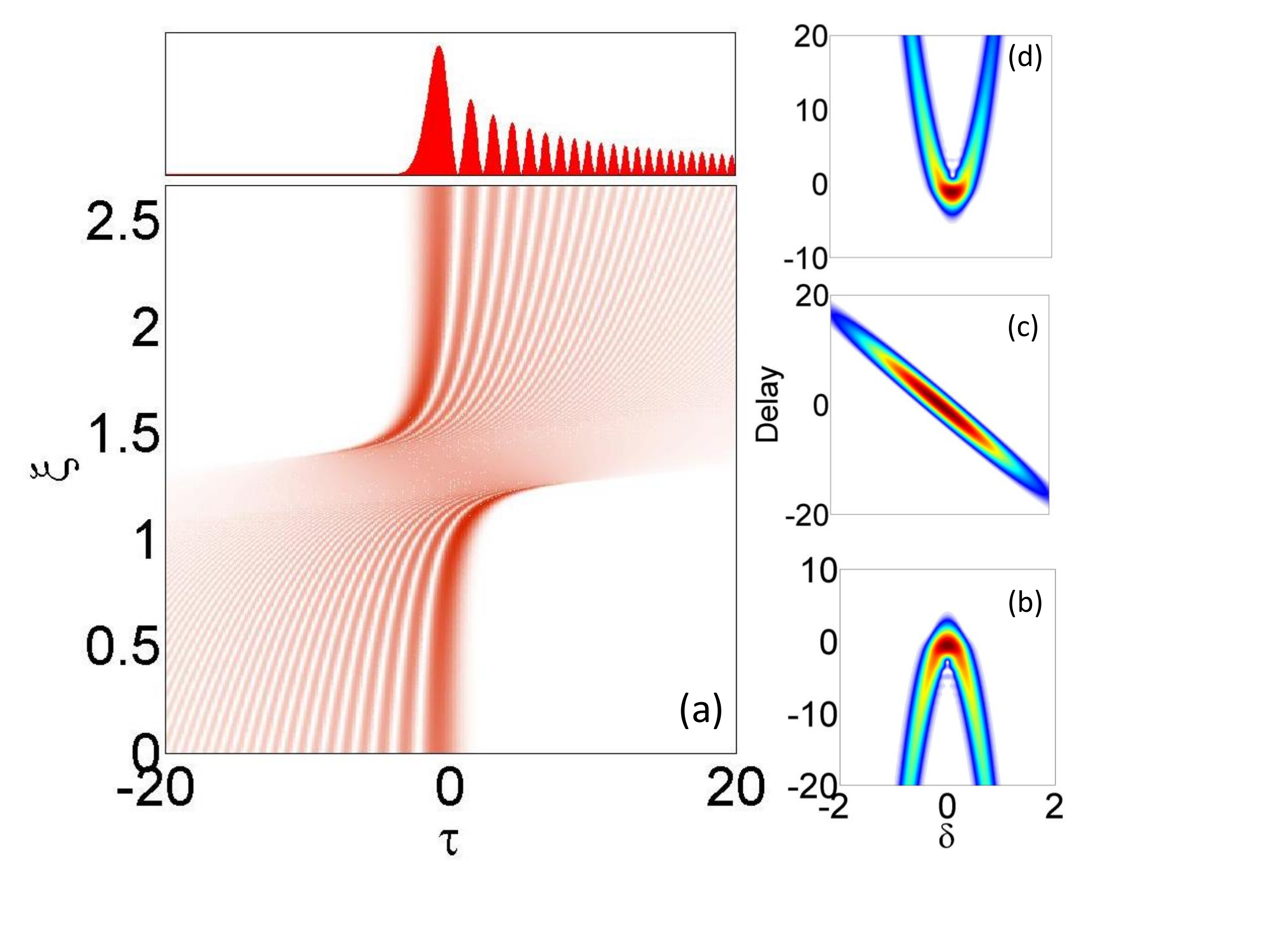,trim=0.2in 0.6in 1.2in 0.0in,clip=true, width=92mm}

\vspace{-2em}
\caption{ (a) Evolution of a FEAP($a=0.01$ and $P_0=1$) under TOD where flipping is observed at $\xi_{flip}\approx1.33$. In the simulation the TOD coefficient is taken as, $\delta_3=0.25$. The upper panel shows the temporarily inverted output. The spectrograms at (b) input, (c) flipping point and (d) output are shown in the right panel.For the spectrograms we have taken $\delta=(\nu-\nu_0)t_0$.Where,$\nu_0$ is the central frequency. }
\label{fig1}
\end{center}
\end{figure}


\noindent Hence the flipping point depends only on $\delta_3$ and for vanishing TOD parameter $\xi_{flip}\rightarrow \infty$. Physically it means, in absence of TOD, there will be no temporal flipping of an Airy pulse. In Fig. \ref{fig1}(a) we show the dynamics of a FEAP under TOD for $\delta_3=0.25$. The temporal flipping is observed at $\xi_{flip}=(3\delta_3)^{-1}\approx1.33$.  With increasing $\delta_3$, the position of flipping and the size of flipping area, both decreases significantly.  In order to grasp the complete dynamics, we adopt cross-correlation frequency resolved optical grating (XFROG) spectrogram technique which plot the frequency and its temporal counter part together. The XFROG is mathematically defined as the convolution $s(\tau,\omega,\xi)=|\int_{-\infty}^{\infty} u(\xi,\tau')u_{ref}(\tau-\tau')\exp(i\omega\tau')d\tau'|^2$ , where $u_{ref}$ is the reference window function normally taken as an input \cite{Agarwal}. The spectrograms at three different lengths are  shown in the right panel Fig. \ref{fig1}(b)-\ref{fig1}(d) where temporal flipping is evident. Interestingly, at flipping point, the Airy pulse is converted to a linearly chirped Gaussian pulse (see Fig. \ref{fig1}(c)). For a very high $\delta_3$ value, the finite flipping area reduces to a point. This property of FEAP is reported earlier by Driben et. al. \cite{driben}. However in the later section we show that, high $\delta_3$ is not the only necessary condition for absolute focusing, rather with suitable PM we can achieve point like focusing more efficiently.

\subsection{Linear phase modulation (LPM) and Quadratic phase modulation (QPM)}

\begin{figure}[h!]
\begin{center}
\epsfig{file=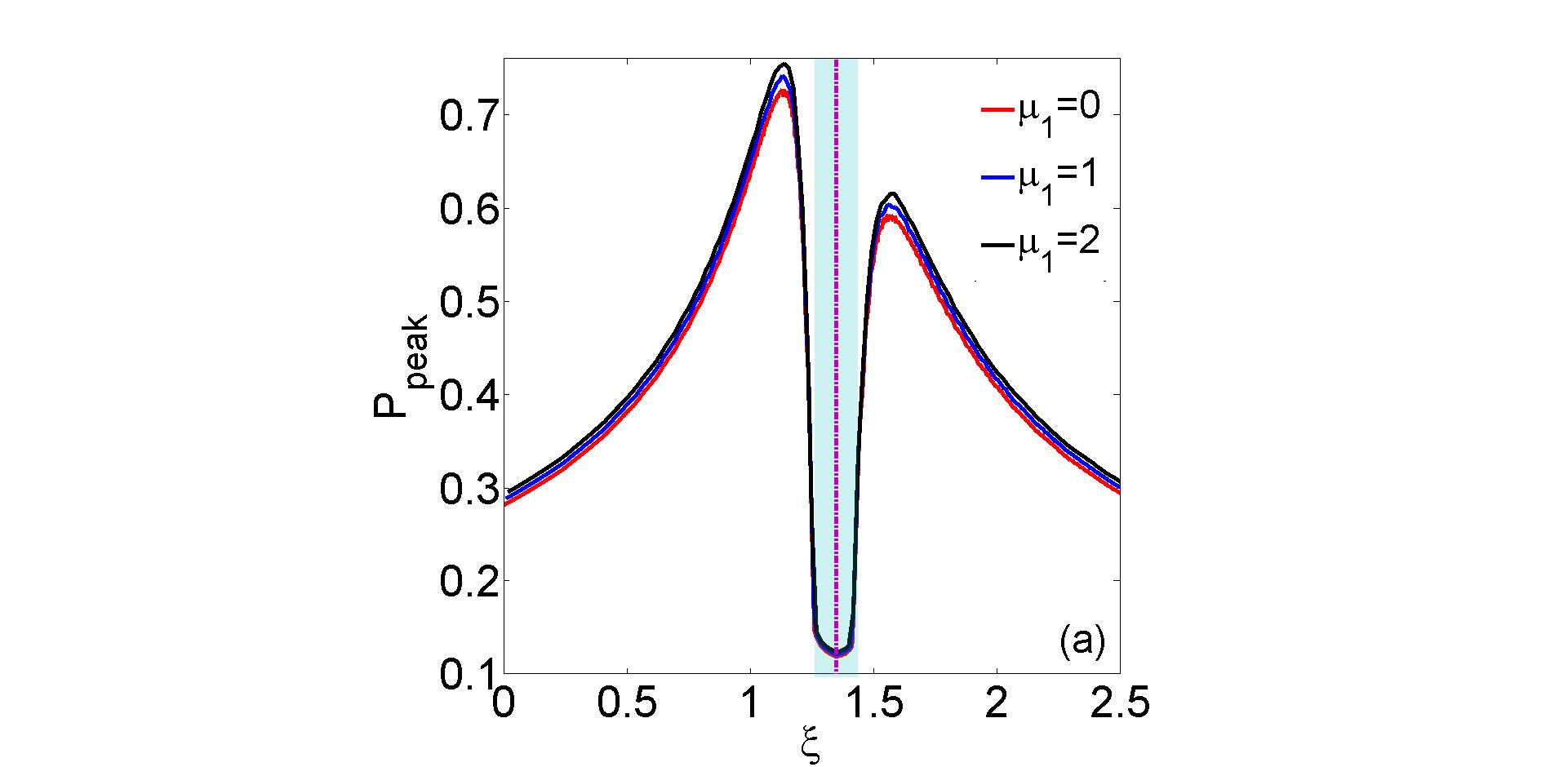,trim=6.0in 0.05in 7.0in 0.0in,clip=true, width=42mm}
\epsfig{file=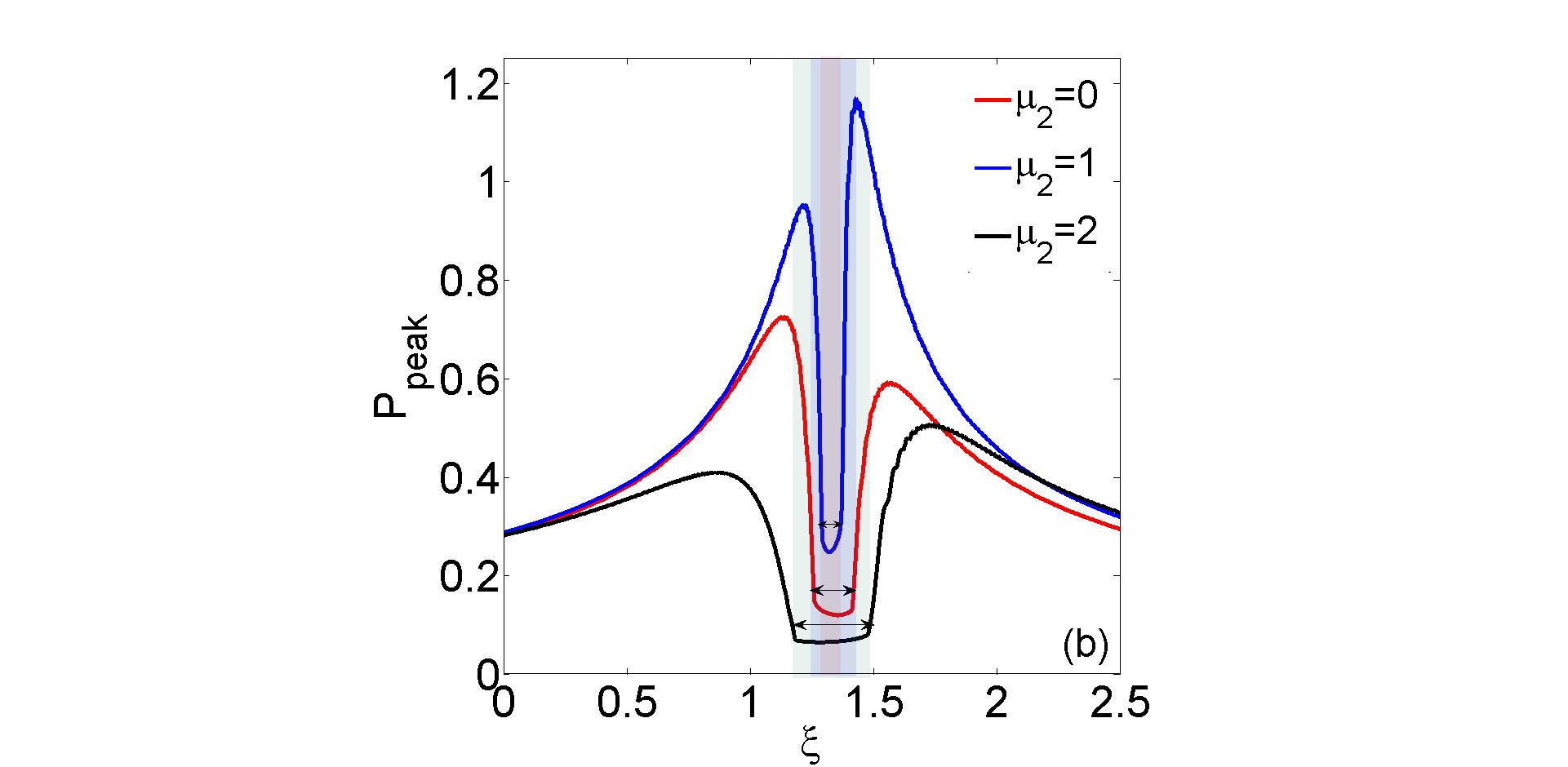,trim=6.0in 0.05in 7.0in 0.0in,clip=true, width=42mm}
\vspace{0em}
\caption{The variation of the peak power of a FEAP having truncation parameter $a$=0.01 for (a) different LPM  $(\mu_1)$. and  (b) QPM parameter $(\mu_2)$. For both the cases TOD coefficient is fixed at $\delta_3=0.25$. Vertical dotted line indicates the location of $\xi_{flip}$. The shades indicate the flipping region defined by the width of the valley. }
\label{fig2}
\end{center}
\end{figure}

\noindent For a linearly phase-modulated input pulse, the solution of Eq. \eqref{q2} can be derived by plugging Eq. \eqref{q5} (with $k=1$) into Eq. \eqref{q3}. The solution becomes,

\begin{equation} \label{q8}
U\left( \xi ,\tau  \right)=\frac{1}{c}\exp \left( \frac{{{a}^{3}}}{3} \right)Ai\left( \frac{{{b}'}}{c}-\frac{{{n}^{2}}}{{{c}^{4}}} \right) \exp i\left( \frac{2{{n}^{3}}}{3{{c}^{6}}}-\frac{n{b}'}{{{c}^{3}}} \right),
\end{equation}
where ${b}'=(b+{{\mu}_{1}})$.  It should be noted that, the condition of singularity is not affected by LPM. It means that ${{\xi }_{flip}}$ should not change under LPM. To confirm it, we have solved the problem numerically and it matches well with our analytical prediction. In Fig.  \ref{fig2}(a) we have plotted the peak power $(P_{peak})$ of the propagating FEAP, without ( $\mu_1=0$) and with ($\mu_1\neq0$) LPM,  as a function of the normalised distance for a fixed $\delta_3$. The flipping point can be clearly seen in the figure where the $P_{peak}$ drops to a minimum value (marked by dotted vertical line ) and then it again increases beyond that point which indicates the temporal reversal.

\noindent It can be observed from Fig. \ref{fig2}(a) that the LPM does not influence the dynamics of the pulse in the context of its flipping nature. The flipping point (indicated by the dotted vertical line) and the flipping area (indicated by the shaded region) both are unaffected under LPM. 
 Next we extend out study for QPM for which the pulse shape at $\xi$ can be derived as, 
\begin{equation} \label{q9}
U\left( \xi ,\tau  \right)=\frac{1}{c}\exp \left( \frac{{{a}^{3}}}{3} \right)Ai\left( \frac{b}{c}-\frac{{{m}^{2}}}{{{c}^{4}}} \right)\exp i\left( \frac{2{{m}^{3}}}{3{{c}^{6}}}-\frac{nb}{{{c}^{3}}} \right),
\end{equation}
where $m=(n+\mu_2)$. Again from the analytical expression we observed that the singularity condition is unaffected under QPM and it is evident in Fig. \ref{fig2}(b). However, unlike LPM, the flipping area is significantly influenced by QPM. The width of the flipping area depends on the argument of the Airy function in Eq. \eqref{q9} which contains the truncation parameter ($a$), TOD coefficient ($\delta_3$) and the QPM parameter ($\mu_{2}$). The interplay between these three parameters decides the width of the flipping area. In the later section we will study this effect elaborately.

\subsection{Cubic phase modulation(CPM)} 
 
In this section we will  show that CPM plays an interesting role in preserving the shape of the Airy pulse travelling in a dispersive medium where $\delta_3\neq0$. Under CPM the shape of the propagating Airy pulse is analytically described as, 

\begin{equation} \label{q10}
U\left(\xi,\tau\right)=\frac{1}{{{c}'}}\\
\exp \left( \frac{{{a}^{3}}}{3} \right)Ai\left(\frac{b}{{{c}'}}-\frac{{{n}^{2}}}{{{{{c}'}}^{4}}} \right)\exp~i\left(\frac{2{{n}^{3}}}{3{{{{c}'}}^{6}}}-\frac{nb}{{{{{c}'}}^{3}}} \right),
\end{equation}\\
where ${c}'={{\left( 1+3{{\mu}_{3}}-3{{\delta }_{3}}\xi  \right)}^{\frac{1}{3}}}$. The peak position of the main lobe of Airy pulse $\tau_p$ is given as,
\begin{equation}\label{q10a}
 \tau_p\approx c'\tau_{p0}+\frac{\xi^2}{4c'^3}+\frac{3a^2}{c'^3}(\mu_3-\delta_3\xi).
\end{equation}
The peak position  now moves at different rate with propagation distance and can be controlled by CPM parameter. It is also interesting to note that, due to the presence of CPM the singularity condition  of the pulse is now modified which leads to a different expression of the flipping point as follows,
\begin{equation}\label{q11}
\xi' {_{flip}}=\frac{1}{3{{\delta }_{3}}}\left( 1+3{{\mu}_{3}} \right).
\end{equation}

 \begin{figure}[h!]
 \begin{center}
 \epsfig{file=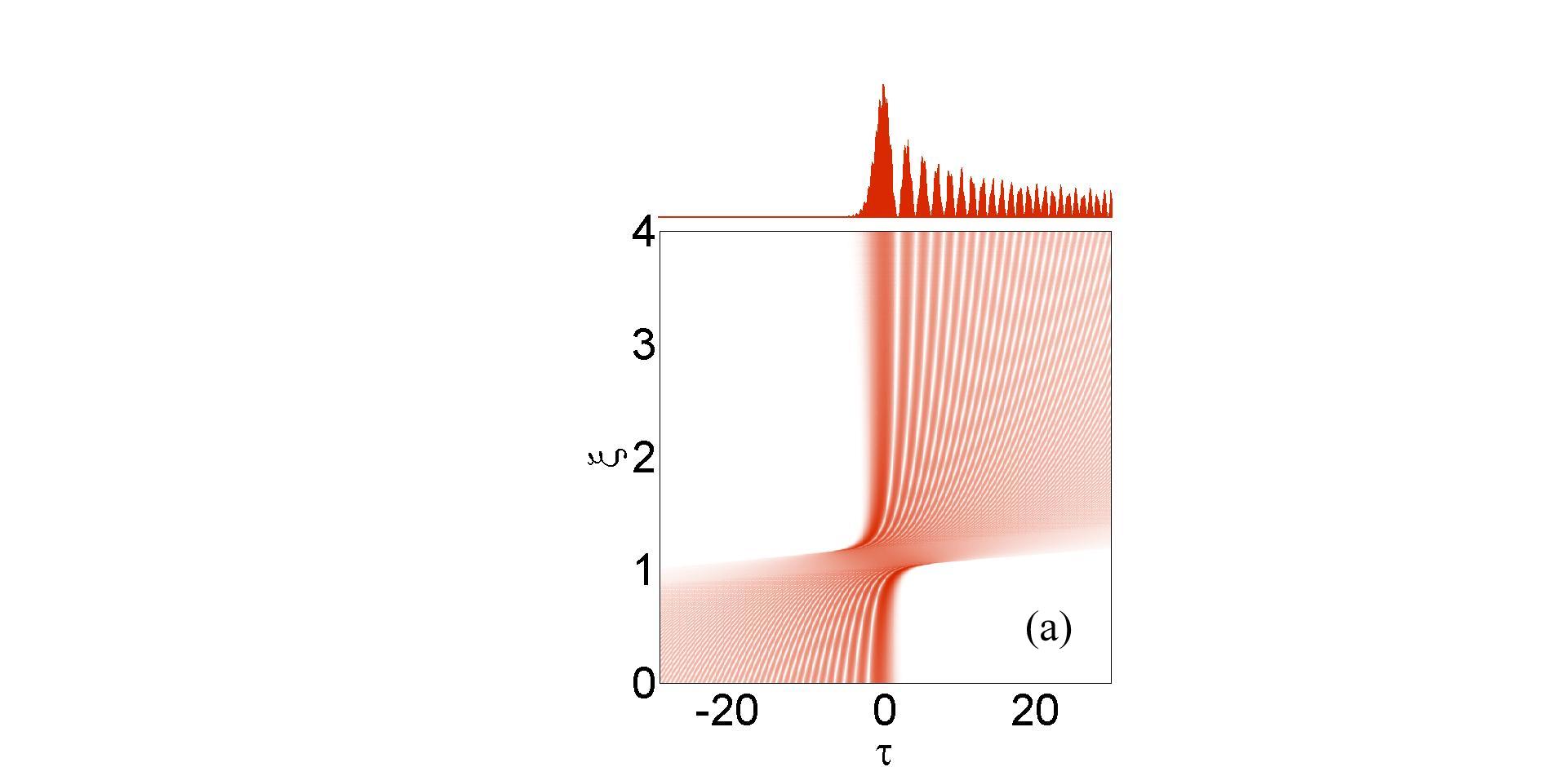,trim=9in 0.0in 7in 0in,clip=true, width=0.5\columnwidth}\epsfig{file=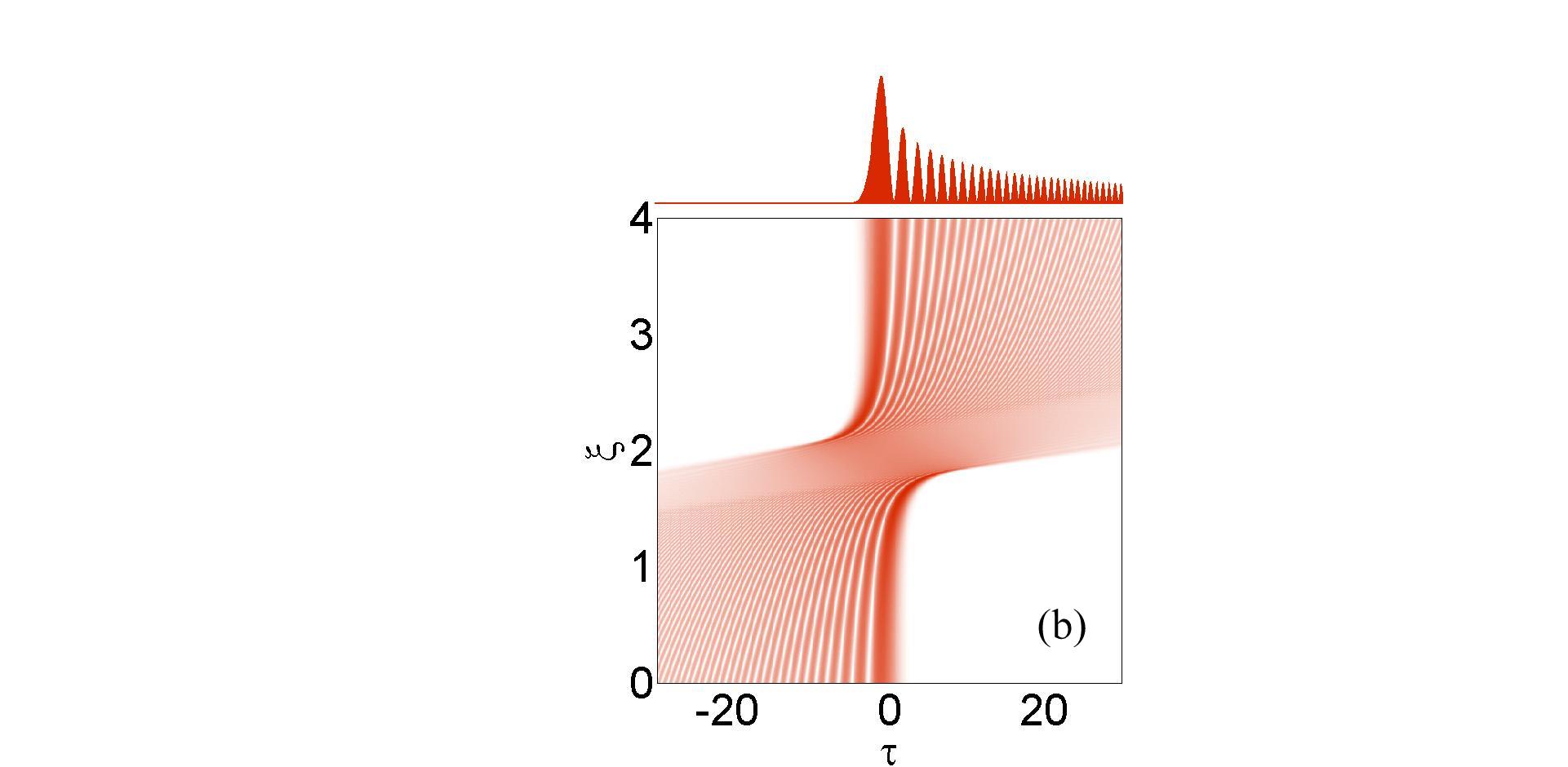,trim=9in 0.0in 7in 0in,clip=true, width=0.5\columnwidth}\vspace{0em}
 \epsfig{file=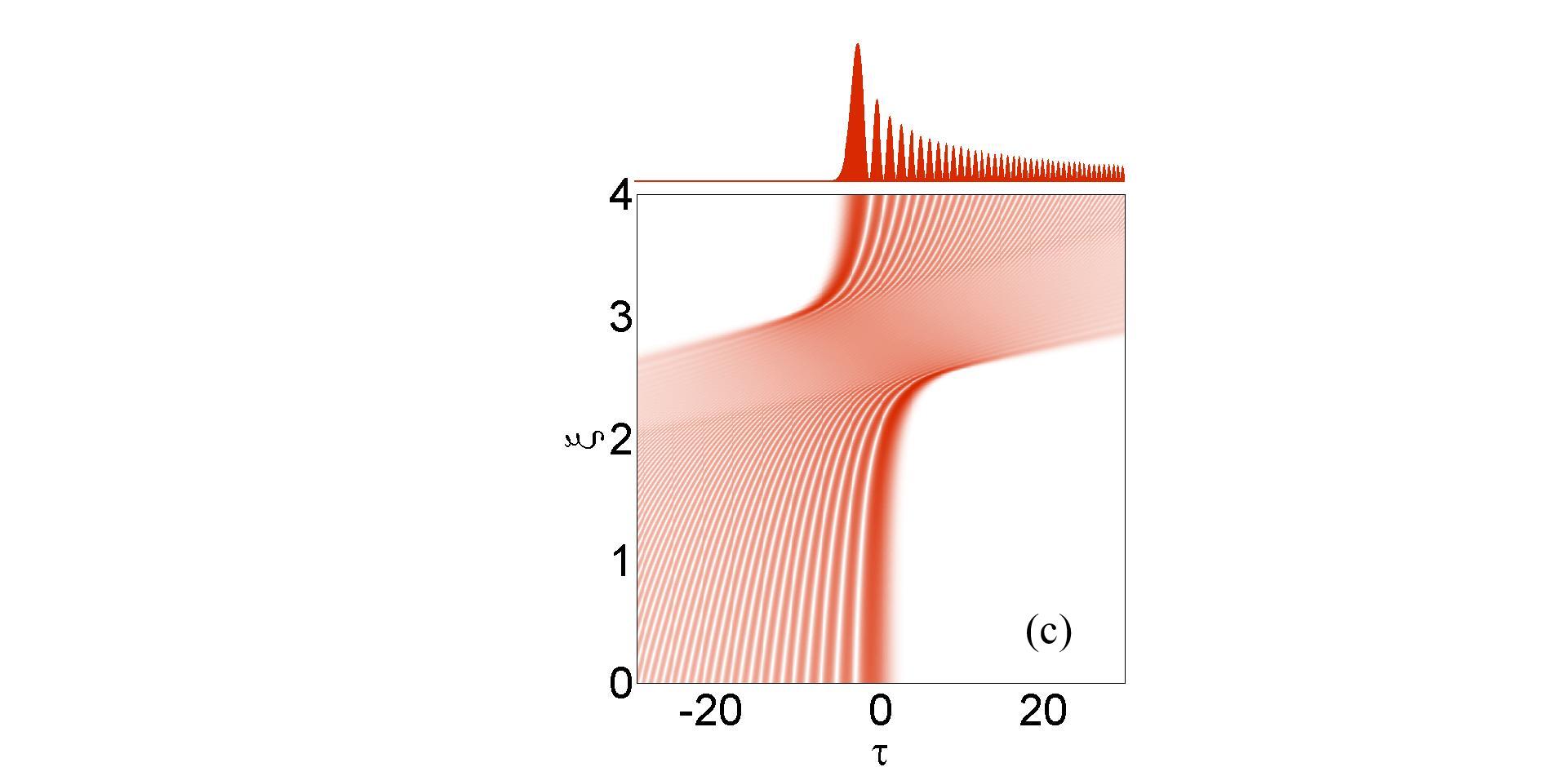,trim=9in 0.0in 7in 0.6in,clip=true, width=0.5\columnwidth}\epsfig{file=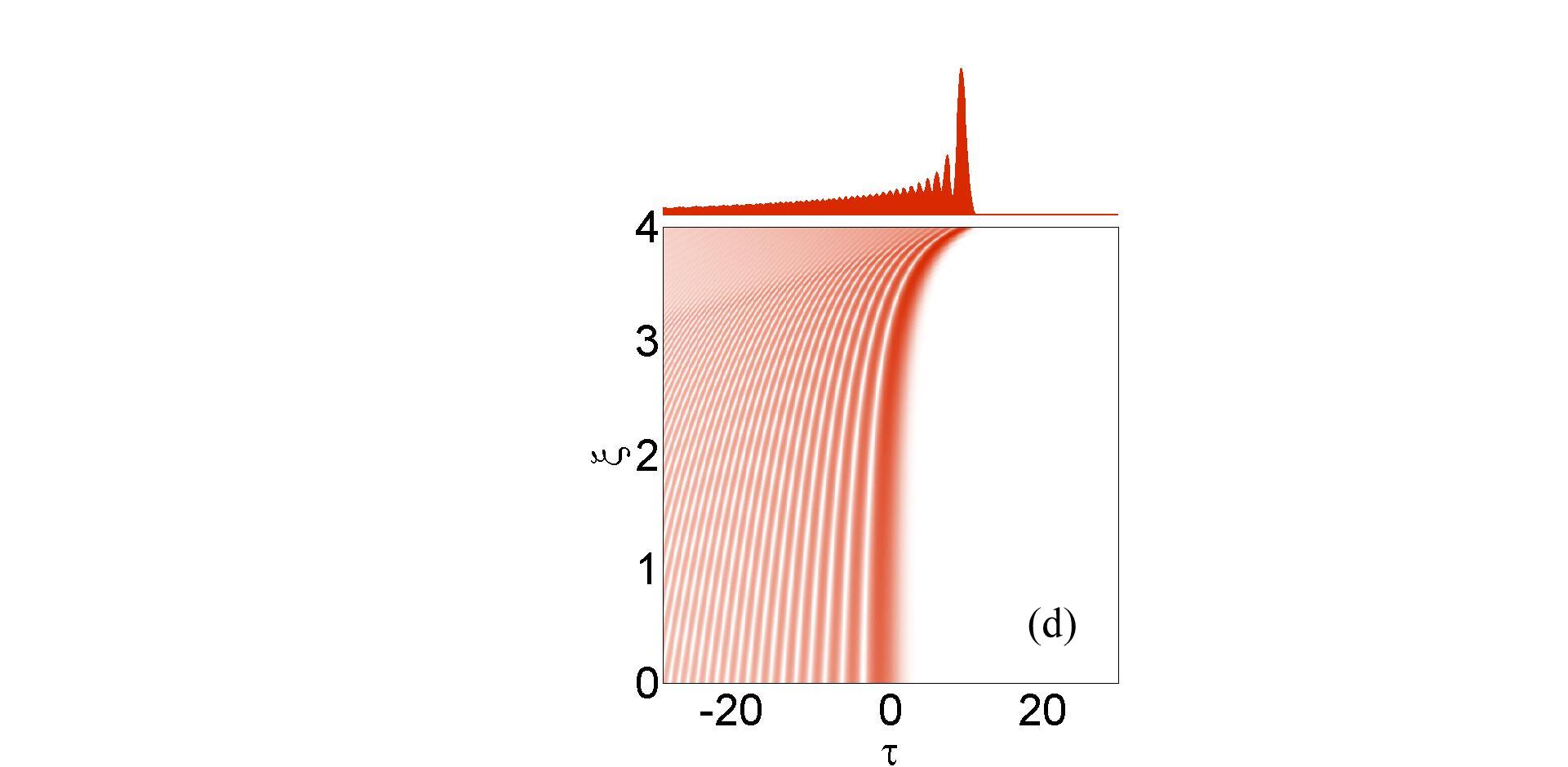,trim=9.5in 0.0in 7in 0.6in,clip=true, width=0.5\columnwidth}\vspace{0em}
 \caption { Propagation dynamics of a truncated Airy pulse in the presence of TOD with $\delta_3=0.3$ for (a) $\mu_3=0$ (b) $\mu_3=0.25$ (c) $\mu_3=0.50$  and (d) $\mu_3=1$. For each case the output shape of the pulse is shown in the upper panel.} 
 \label{fig3}
 \end{center}
 \end{figure}
 
\begin{figure}[h!]

\begin{center}
\epsfig{file=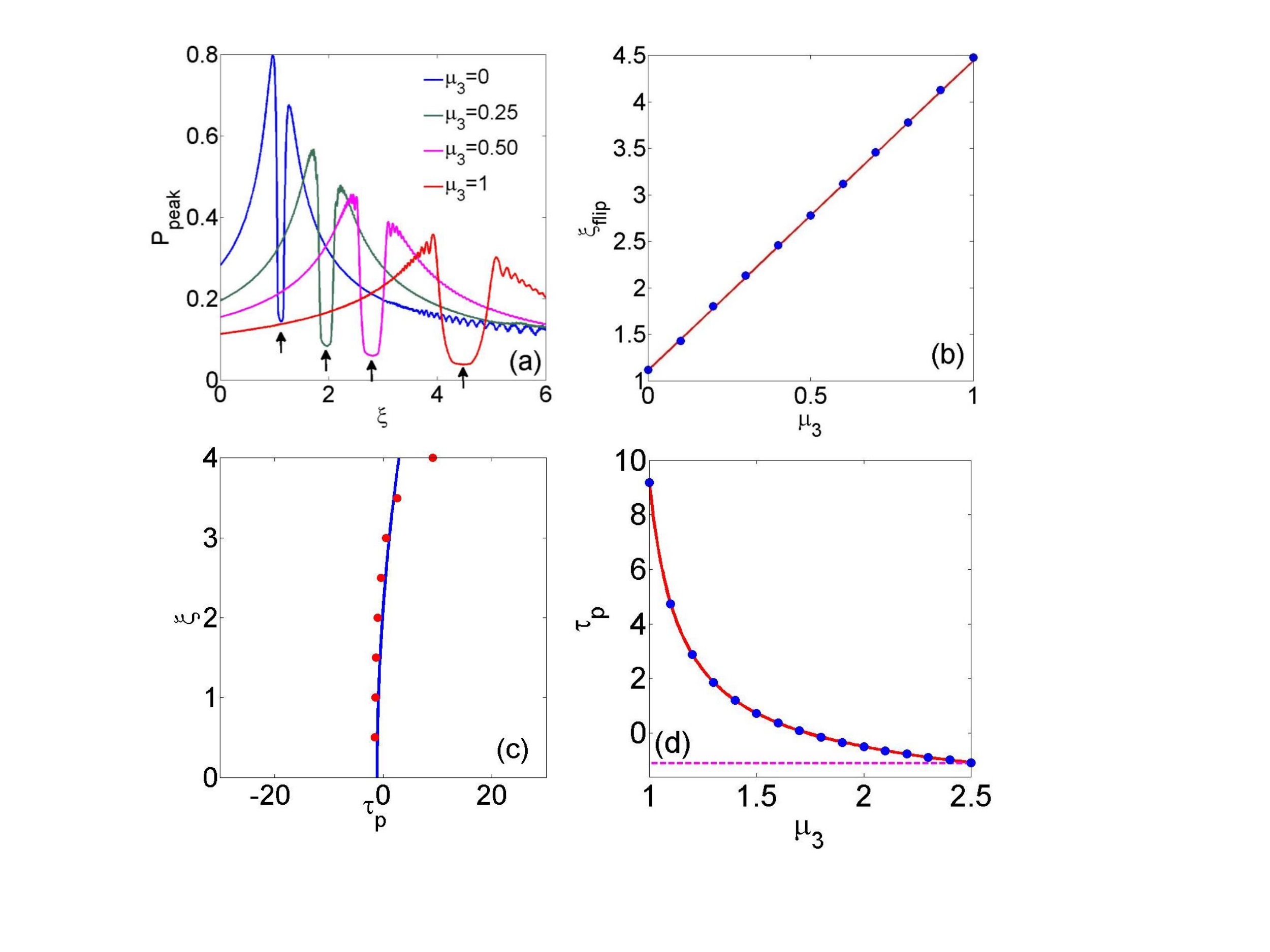,trim=1.1in 0.8in 1.5in 0.1in,clip=true, width=92mm}

\vspace{0em}
\caption{ (a) Variation of peak intensity of the propagating pulse with $\xi$ for several CPM parameters ($\mu_3$) when $\delta_3=0.3$. The arrows indicate the exact position of flipping ($\xi_{flip}$). (b)  Plot of $\xi_{flip}$ as a function of $\mu_3$. (c) The natural trajectory of the Airy peak (solid line) in absence of TOD is compared with the phase modulated Airy peak (dots) with $\delta_3=0.3$ and $\mu_3=1$. (d) Variation of $\tau_p$ with $\mu_3$ at a fixed distance $\xi=4$ when $\delta_3=0.3$. The dotted line indicates the peak position ($\tau_{p0}$) of Airy pulse at input. In plot (b) and (d), the solid line and dots indicate the theoretical and numerical data, respectively. } 
\label{fig4}
\end{center}
\end{figure}

 
This new expression suggests that we can increase the flipping distance by increasing the  CMP parameter (positive values) and  can in principle preserve the pulse shape for a longer distance. This is an interesting aspect of CPM which was never addressed before. In Fig. \ref{fig3} we show the evolution of the truncated Airy pulse for different CMP parameters ($\mu_3$) with a fixed $\delta_3$. From the simulated figures it is evident that, the temporal flipping due to $\delta_3$ can be counter balanced by the CMP parameter $\mu_3$ so that the Airy pulse retains its shape at output. In addition to the temporal flipping, TOD distorts the overall pulse shape  (see upper panel of Fig. \ref{fig3}(a)). We demonstrate that CPM can also prevent this distortion (Fig. \ref{fig3}(d)). At the flipping region Airy pulse loses its characteristics and becomes a pure Gaussian at $\xi_{flip}$. A rapid decrement of peak power is observed at $\xi_{flip}$. In Fig. \ref{fig4}(a) we show the variation of peak power as a function of distance for different CPM parameters. The exact values of $\xi_{flip}$ can be evaluated by the locations of dips (indicated by arrows). It should be noted that, with increasing $\mu_3$ the flipping region (defined as the width of the valley) and the  distance of the flipping point both increases. In Fig. \ref{fig4}(b) we plot $\xi_{flip}$ as a function of $\mu_3$ which shows a linear relationship, i.e, with increasing $\mu_3$ the flipping point increases linearly. We investigate this feature numerically and  the numerical data (solid dots) corroborate well with theoretical prediction given in Eq. \eqref{q11}  (solid line). In Fig. \ref{fig4}(c) the solid line represent the natural trajectory of the Airy peak when  TOD and phase modulation both are absent. In the same plot, using dots we superimposed the path followed by the Airy peak when $\delta_3=0.3$ and $\mu_3=1$. The paths indicated by dots and solid line are found very close to each other. Hence a suitable PM can counter balance the perturbing effect of TOD and preserve the ideal trajectory. In Fig. \ref{fig4}(d) we show how the CPM parameter $\mu_3$ can tailor the peak position at some fixed distance (here $\xi=4$). The horizontal dotted line indicates $\tau_{p0}$    which is the peak location of Airy pulse at $\xi=0$. Hence suitable $\mu_3$ not only preserve the pulse shape but for a given distance can produce a nearly bend free Airy pulse without compromising its self-healing property.
 
\subsection{Effect of PM on the flipping area}
  
Under the perturbation of TOD, Airy pulse exhibits unusual dynamics when it flips in temporal domain. The Airy pulse loses it characteristics and converges to a pure Gaussian shape at the singular point. In general, there is a finite region where the Airy pulse loses its shape and the singular point is typically located in the middle of this region.     We call this region as \textit{flipping region} which is characterised by the width of the Gaussian pulse. We investigate that, the length of this flipping region is proportional to the width of the Gaussian pulse obtained at the flipping point. 

 \begin{figure}[h!]
  \begin{center}
  \epsfig{file=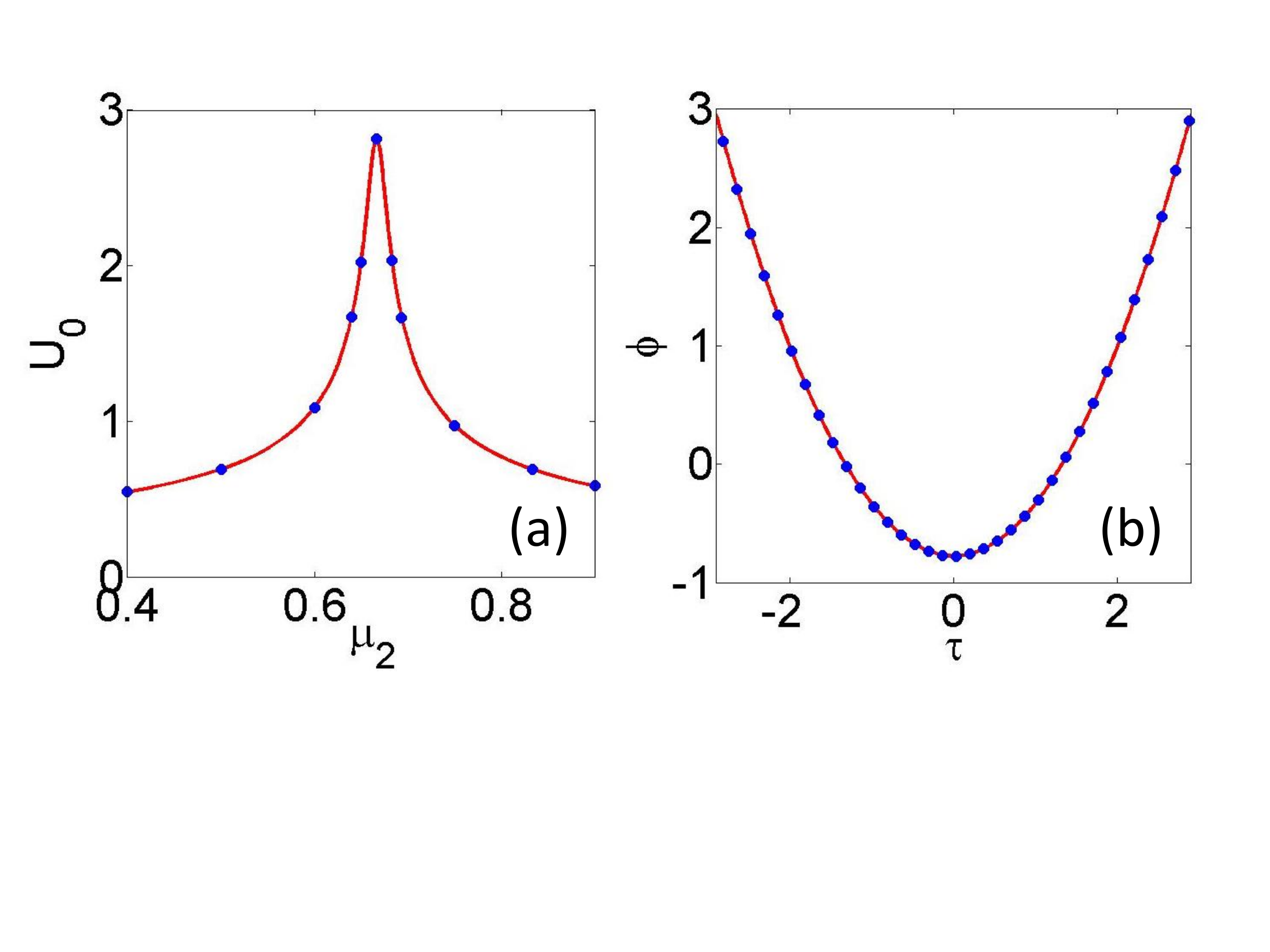,trim=0.2in 2.2in 0.5in 0.7in,clip=true, width=75mm}
  
  \caption{(a)The amplitude ($U_0$) of the gaussian pulse as a function $\mu_2$ for a fixed $\delta_3$ value of 0.25. (b) the variation of the phase ($\phi$) of the Gaussian pulse for a fixed QPM parameter ($\mu_2=0.1$)} 
  \label{fig5}
  \end{center}
  \end{figure}

In the previous work, Driben et. al \cite{driben} qualitatively demonstrated that, the flipping region can be squeezed to a point by launching an Airy pulse close to zero dispersion frequency where TOD parameter is very high. In our study, however, we find that QPM can play a dominant role in controlling the flipping region. In fact, even for a finite $\delta_3$ the flipping region reduces to a point for some critical value of $\mu_2$. In presence of QPM, the solution at singular point can simply be expressed as, 
   
\begin{equation}\label{q12}
U(\xi_{flip},\tau)=U_0 \exp \left[-\frac{b^2}{\tau_{f}^2} \right]\exp(i\phi),
\end{equation}
where the characteristic width of the Gaussian pulse is, $\tau_f=2\sqrt{a(1+\Delta^2/a^2)}$ and  $\gamma =\sqrt{(a^2+\Delta^2)} $. The detuned parameter $\Delta$ is defined as, $\Delta=[\mu_2-(6\delta_3)^{-1}]$. The amplitude ($U_0$) and phase ($\phi$) of the Gaussian pulse at flipping point are calculated as,
\begin{equation}\label{q12a}
U_0= \frac{1}{2\sqrt{\pi\gamma}}\exp(a^3/3), 
\end{equation}
and
\begin{equation}\label{q12b}  
\phi=\frac{1}{2}\tan^{-1}\left(\frac{\Delta}{a}\right)-\frac{\Delta(\tau-a^2)^2}{4\gamma^2}.   
\end{equation}  
 It is interesting to note that, at zero detuning ($\Delta=0$), the peak amplitude of the Gaussian pulse reaches to a maxima and the total phase of the pulse becomes zero, we define it as \textit{absolute focusing}. We study the behaviour of the Gaussian pulse in the flipping region in detail. In Fig. \ref{fig5}(a) we plot the variation of the peak amplitude of the Gaussian pulse at $\xi_{flip}$ as a function of $\mu_2$. At absolute focusing the peak of the Gaussian pulse reaches to a maxima and this is achieved when $\mu_2=(6\delta_3)^{-1}$, in this case it is around 0.667 since $\delta_3$ is fixed at 0.25. In the figure the solid line corresponds to the analytical expression of peak amplitude where as, dots represents the results obtained from direct simulation. The phase of the Gaussian pulse is also plotted in  Fig. \ref{fig5}(b) where the value of $\mu_2$ is kept fixed at 0.1. The analytical curve (solid line) corroborate well with the numerical solution (solid dots).

     
\begin{figure}[h!]
\begin{center}

\epsfig{file=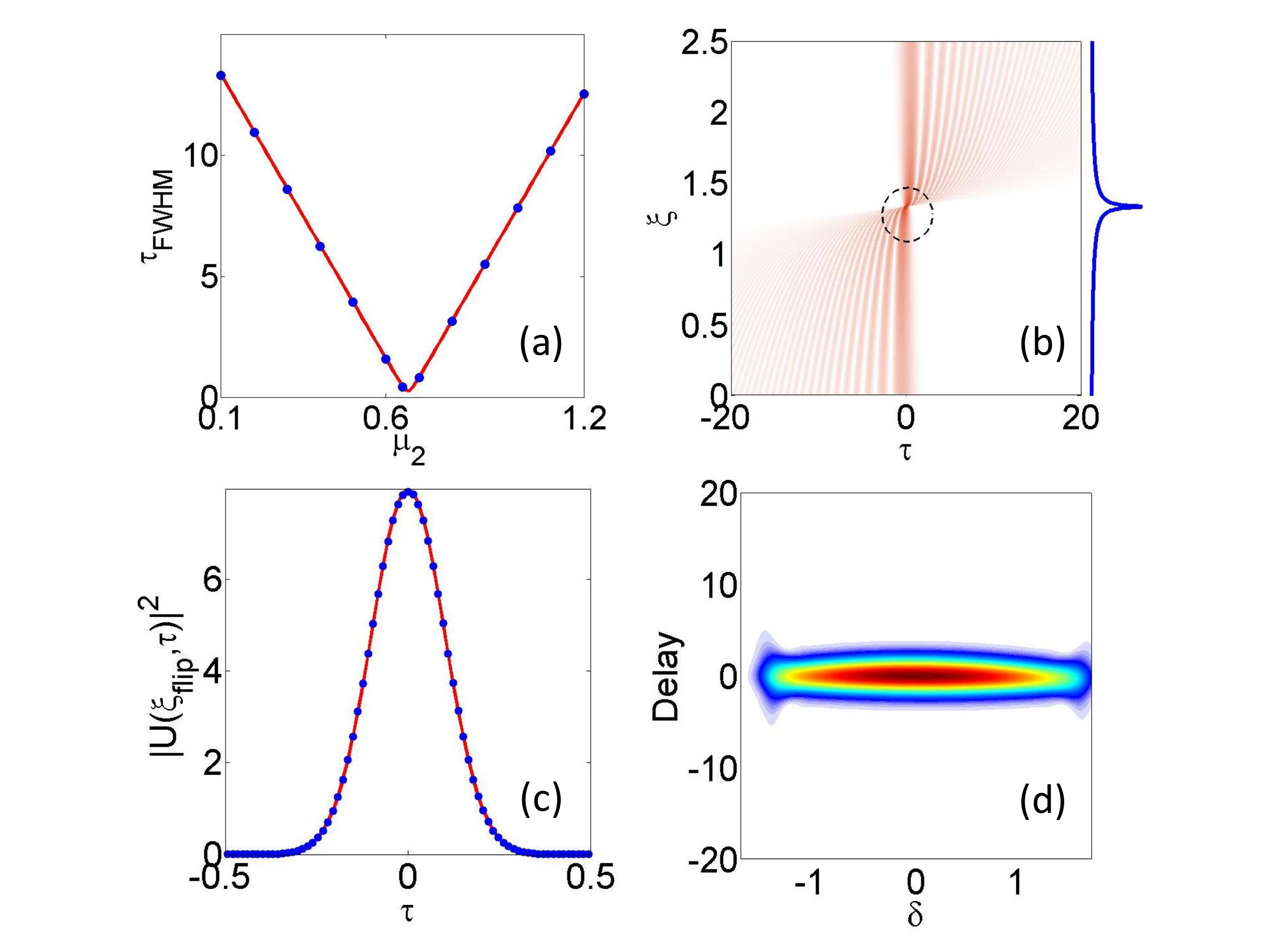,trim=1in 0.0in 1in 0.2in,clip=true, width=80mm}
\vspace{-1em}
\caption{(a)Variation of $\tau_{FWHM}$ ($=\tau_{f}\sqrt{2ln2}$) as a function of $\mu_2$ , for $\delta_3=0.25$. The solid dots indicate the widths calculated numerically where as solid line corresponds the analytical expression . (b) Evolution of a FEAP for a fixed $\delta_3=0.25$ with $\mu_2=(6\delta_3)^{-1}$. The dotted circle indicates the location of absolute focusing. In the right panel the evolution of the peak power over $\xi$ is shown. (c) The intensity distribution of the confined gaussian pulse obtained  the point of absolute focussing.The analytical expression (red line) obtained in Eq. \ref{q12}  corroborates well with the numerically found solution (blue dots). (d) The XFROG spectrogram in the case of absolute focussing ($\mu_2=(6\delta_3)^{-1}$).}
\label{fig6}
\end{center}
\end{figure}

In Fig. \ref{fig6}(a) we plot the $\tau_{FWHM}$ 
as a function of $\mu_2$ (for $\delta_3=0.25$), where we achieve a tight focusing at $\mu_2\approx0.67$, as predicted by the theory. Here the dots represent the values of FWHM obtained numerically and the analytical result is shown by solid line. In Fig. \ref{fig6}(b) we show the evolution of a FEAP where $\delta_3=0.25$ and  $\mu_2=(6\delta_3)^{-1}$. In the right panel of the Fig. \ref{fig6}(b) it is shown how $P_{peak}$ reaches to a maxima (at $\xi_{flip}$) when absolute focusing is achieved.
In Fig. \ref{fig6}(c) we have plotted the intensity distribution of the Gaussian pulse obtained at $\xi_{flip}$ (Eq.\ref{q12}) when the condition of absolute focussing is satisfied. The numerical solution (blue solid dots) agrees well with analytical expression  Eq.\eqref{q12}. In order to grasp the total picture at absolute focusing, we examine the XFROG spectrogram as shown in Fig. \ref{fig6}(d). The spectrogram clearly indicates the  tight temporal confinement of the pulse with zero chirp.

\begin{figure}[h!]
  \begin{center}
  \epsfig{file=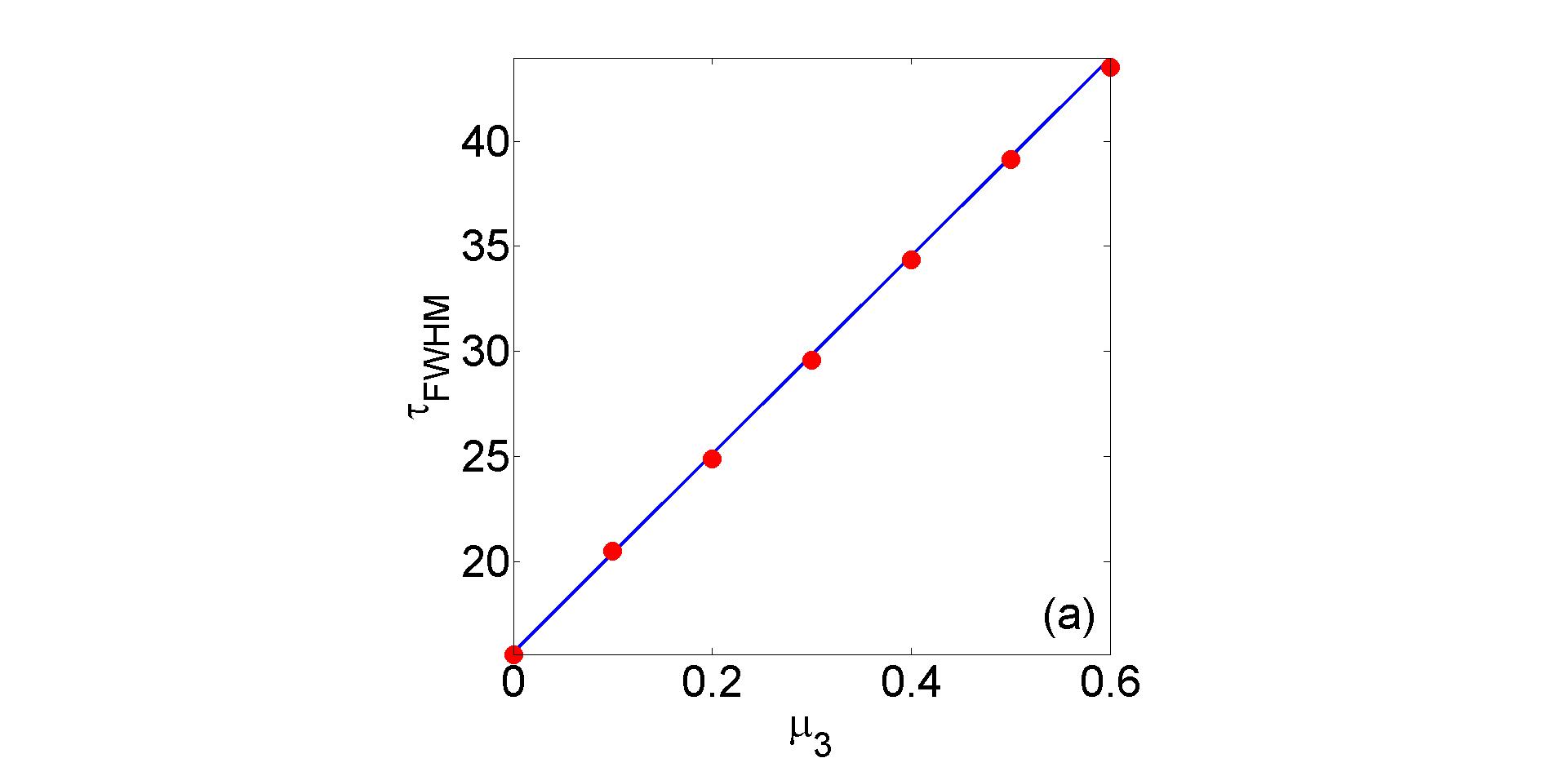,trim=6.6in 0.05in 7.0in 0.0in,clip=true, width=42mm}
  \epsfig{file=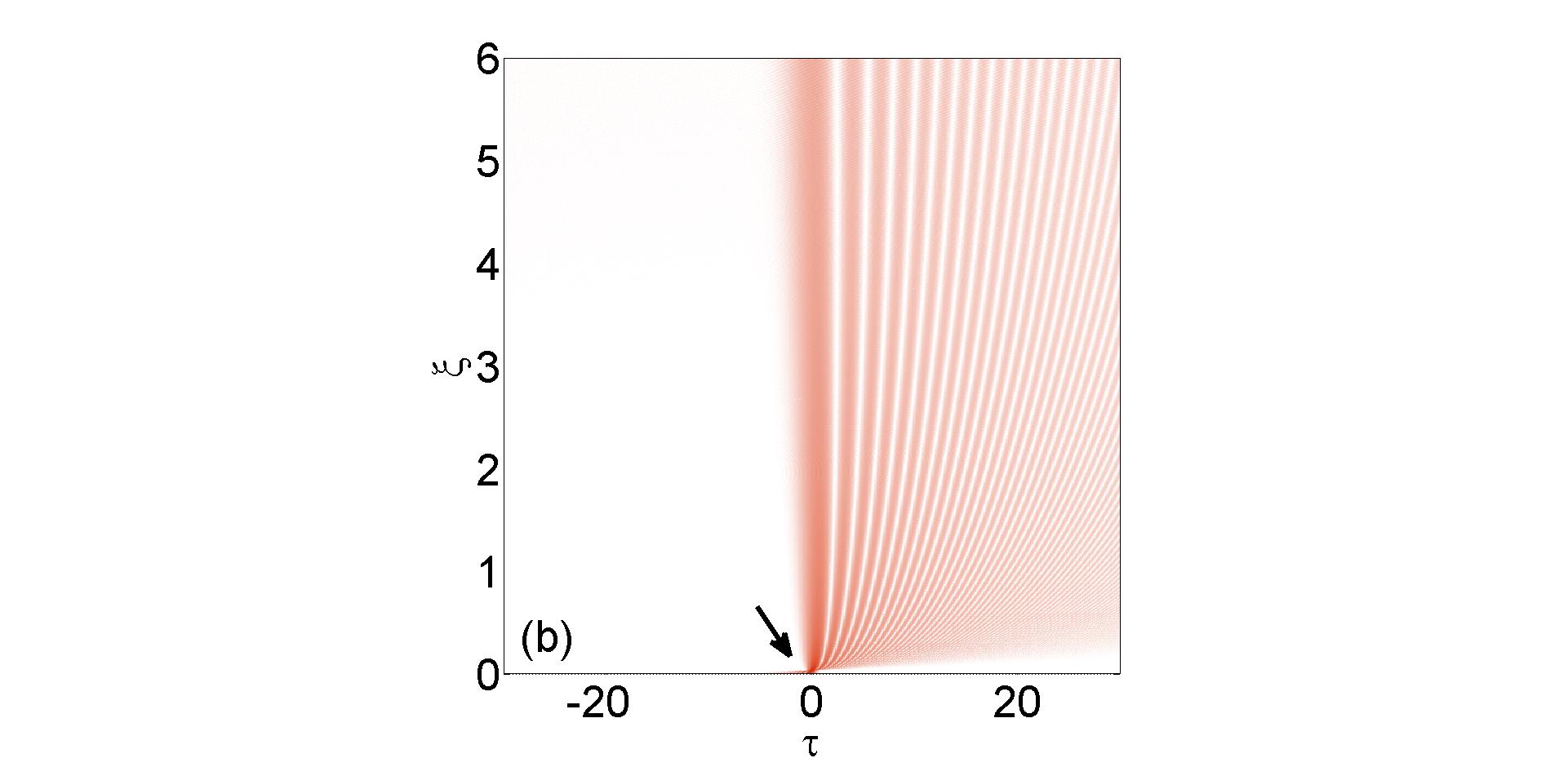,trim=6.6in 0.05in 7.0in 0.0in,clip=true, width=42mm}
  \vspace{0em}
  \caption{ (a)Gaussian pulse width as a function of CPM parameter $\mu_3$ when $(\delta_3=0.25)$. (b)Evolution of the Airy pulse under critical $\mu_3$ when absolute focusing is occurred at $\xi \approx0$ point (indicated by the arrow).}
  \label{fig7}
  \end{center}
  \end{figure}

 CPM can also contribute in modifying the width of the Gaussian pulse that appears at the flipping region. For CPM, the expression of the width of that Gaussian pulse at $\xi_{flip}$ is derived as, 
 
\begin{equation}\label{q14}
{{\tau }_{FWHM}}=2\sqrt{2\ln2} {{\left[ a+\frac{{{\left( 1+3{{\mu }_{3}} \right)}^{2}}}{36a\delta _{3}^{2}} \right]}^{\frac{1}{2}}}
\end{equation}
Form the expression in Eq. \eqref{q14} it is clear that, for $\mu_3>0$, width will increase monotonically. Since $a$ is small we can approximate the expression as, $\tau_{FWHM}\approx \tau_b(1+3\mu_3)$, where $\tau_b=(3\delta_3)^{-1}\sqrt{2\ln2/a}$. Hence the width of the Gaussian pulse increases linearly with $\mu_3$. In Fig. \ref{fig7}(a) we plot the evolution of the width (at flipping point) as a function of $\mu_3$ where analytical expression (solid line) corroborate well with the simulated data (dots).  Note that, under CPM we may have absolute focusing when $\mu_3\rightarrow -1/3$. However, in such case $\xi^{'}_{flip} \rightarrow 0$. To visualise this condition in Fig. \ref{fig7}(b) we plot the evolution of a FEAP for $\mu_3\approx-1/3$. In the figure the absolute focusing point is indicated by the arrow. It is obvious that when $\mu_3$ is very close to the critical value $-1/3$,  the Airy pulse will invert almost at the input and propagate without any distortion.


\section{conclusion}
We study the role of linear, quadratic and cubic phase modulation on Airy pulse dynamics. In specific, cubic phase modulation (CPM) can influence the natural parabolic trajectory of an Airy pulse and can assist the pulse to maintain the shape during its motion in a dispersive medium for a longer propagation distance. Under the perturbation of third order dispersion (TOD) a finite energy Airy pulse is inverted temporally and propagates with a reverse acceleration. At the flipping point, the propagating Airy pulse loses its characteristic shape and forms a pure Gaussian structure. We show, in particular, cubic phase modulation acts as a healing effect by preserving the pulse shape during propagation when TOD is on.  CPM counter-balance the flipping effect initiated by TOD. The flipping point can be tailored desirably by using suitable CPM parameter. We also study the role of linear phase modulation(LPM) and quadratic phase modulation (QPM) on pulse dynamics and observe that QMP can play a dominate role in focusing the pulse tightly at flipping point. All analytical results are verified with direct simulation and we find a satisfactory agreement. Our study unfolds few useful and interesting aspects of the dynamics of a phase-modulated Airy pulse, that was unexplored before . 
\section*{ACKNOWLEDGEMENTS}
The author A.B. acknowledges MHRD, India for his research fellowship.

\end{document}